\documentclass[a4paper,11pt]{article}
\pdfoutput=1
\usepackage{jheppub}

\usepackage{amssymb,amsmath,amsfonts}
\usepackage[normalem]{ulem}
\usepackage[utf8x]{inputenc}
\usepackage{slashed}
\usepackage{graphicx}
\usepackage{tabularx}
\usepackage{here}
\usepackage{color}
\usepackage{csquotes} 
\usepackage{comment}
\usepackage{mathrsfs}
\usepackage{float}
\usepackage{ascmac}
\usepackage{multirow}
\usepackage{longtable}
\usepackage{caption}
\usepackage{bm}
\usepackage{ulem}
\usepackage{tikz}
\usepackage{tikz-feynman}
\tikzfeynmanset{compat=1.1.0, warn luatex=false}
\usetikzlibrary{decorations.pathmorphing,positioning,shapes.geometric,calc,arrows.meta}

\usepackage{colortbl} 

\usepackage[italicdiff]{physics}
\makeatletter
\newcommand*\rel@kern[1]{\kern#1\dimexpr\macc@kerna}
\newcommand*\widebar[1]{%
  \begingroup
  \def\mathaccent##1##2{%
    \rel@kern{0.8}%
    \overline{\rel@kern{-0.8}\macc@nucleus\rel@kern{0.2}}%
    \rel@kern{-0.2}%
  }%
  \macc@depth\@ne
  \let\math@bgroup\@empty \let\math@egroup\macc@set@skewchar
  \mathsurround\z@ \frozen@everymath{\mathgroup\macc@group\relax}%
  \macc@set@skewchar\relax
  \let\mathaccentV\macc@nested@a
  \macc@nested@a\relax111{#1}%
  \endgroup
}
\makeatother

\numberwithin{equation}{section}

\preprint{
\begin{minipage}{5cm}
\small
\flushright
EPHOU-25-014\\
KYUSHU-HET-334
\end{minipage}}

\title{Non-invertible Symmetry as a Solution to the Strong CP Problem in a GUT-inspired Standard Model}

\author{Tatsuo Kobayashi$^{1}$,} 
\author{Hajime Otsuka$^{2}$, and}
\author{Tsutomu T. Yanagida$^{3,4}$}
\affiliation{
$^1$Department of Physics, Hokkaido University, Sapporo 060-0810, Japan\\
$^2$Department of Physics, Kyushu University, 744 Motooka, Nishi-ku, Fukuoka 819-0395, Japan\\
$^3$Kavli IPMU (WPI), UTIAS, University of Tokyo, Kashiwa, 277-8583, Japan\\
$^4$Tsung-Dao Lee Institute \& School of Physics and Astronomy, Shanghai Jiao Tong University, China
}
\emailAdd{kobayashi@particle.sci.hokudai.ac.jp}
\emailAdd{otsuka.hajime@phys.kyushu-u.ac.jp}
\emailAdd{tsutomu.tyanagida@sjtu.edu.cn}

\abstract{
We propose a three-zero texture for the down-quark mass matrix within a $SU(5)$ GUT-inspired Standard Model, enforced by a non-invertible selection rule originating from $\mathbb{Z}_2$ gauging of $\mathbb{Z}_N$ symmetries. Assuming CP invariance at the high-energy scale, our framework solves the strong CP problem while reproducing a realistic quark mass matrix. The CP symmetry is spontaneously broken down at an intermediate scale via complex vacuum expectation values of new scalar fields $\eta_i$, which generate the CP phase in the CKM matrix without inducing the QCD $\bar{\theta}$ term. The present model incorporates three generations of matter, Higgs, and additional scalars with a common structure under the non-invertible symmetry, which may be naturally embedded into $SO(10)$ GUTs at high energies.
}

\makeatletter
\gdef\@fpheader{}
\makeatother

\begin{document}

\maketitle

\section{Introduction}
\label{sec:Intro}

Symmetries are an important principle in physics.
In particular, symmetries are very important to determine allowed interactions and couplings among the elementary particles, that is, they provide us with consistent interactions  and coupling selection rules. Therefore, group-theoretical symmetries were used in particle physics for a long time.
However, the concept of symmetries has been, recently, generalized including non-invertible symmetries.
(See Refs.~\cite{Schafer-Nameki:2023jdn,Shao:2023gho} for reviews of non-invertible symmetries.)

Non-invertible symmetries were applied in particle physics in these years \cite{Choi:2022jqy,Cordova:2022fhg,Cordova:2022ieu,Cordova:2024ypu}.
Among new symmetries, $\mathbb{Z}_2$ gauging of $\mathbb{Z}_N$ symmetry is very interesting  \cite{Kobayashi:2024yqq,Kobayashi:2024cvp,Funakoshi:2024uvy}.
This is because the new type of symmetry has opened a window in the particle physics to control the mass-matrix textures for quarks and leptons, which can not be realized by the normal group-theory based symmetries \cite{Kobayashi:2024cvp,Kobayashi:2025znw,Kobayashi:2025ldi}.\footnote{See Refs.~\cite{Suzuki:2025oov,Kobayashi:2025cwx,Kobayashi:2025lar,Nomura:2025yoa,Chen:2025awz,Okada:2025kfm} for other applications.}
In particular, it has been recently pointed out that the strong CP problem in QCD can be solved owing to the non-inverted symmetry \cite{Liang:2025dkm} based on a CP invariance in the high-energy fundamental theory. In this letter we extend this LY approach \cite{Liang:2025dkm} in a $SU(5)$-GUT inspired standard model (SM).

The CP problem in QCD is the discrepancy between the theoretically predicted $\mathcal{O}(1)$ CP-violating vacuum angle ${\bar \theta}$ and the experimental upper bound of $ {\bar \theta}<\mathcal{O}(10^{-10})$ from electric dipole moment measurements of the neutron~\cite{Abel:2020pzs}. The CP invariance is the most straightforward solution to the strong CP problem. However, CP invariance must be spontaneously broken down at some intermediate scale to generate complex mass matrices for quarks in order to explain the CP violation in the CKM matrix at the weak scale. The diagonalization of quark mass matrices induces a shift in the physical vacuum angle ${\bar\theta}$ away from zero unless the determinant of mass matrices for the up- and down-type quark, $\det[M_u M_d]$, is real. Thus, the strong CP problem arises in QCD even if CP is an exact symmetry at the fundamental level. 

String theory is a promising candidate for fundamental theory and it has CP symmetry \cite{Green:1987mn}.
One possible scenario for spontaneous CP violation is moduli stabilization through compactifications.
However, simple moduli stabilization leads to CP invariant vacua. 
In particular, if the theory has more symmetries, e.g. modular symmetry in addition to the CP symmetry, those symmetries tend to protect the CP symmetry from its violation. (See e.g. Refs.~\cite{Kobayashi:2019uyt,Kobayashi:2020uaj,Ishiguro:2020nuf,Novichkov:2022wvg,Knapp-Perez:2023nty,Higaki:2024pql}.)
Therefore, an alternative scenario is that CP is spontaneously broken down within the framework of four-dimensional (4D) effective field theory.
Furthermore, we need a proper way for the spontaneous CP violation such that the CP violation appears in the weak CP phase, but not 
in $\det[M_u M_d]$.

The LY model is a proposal to address the above problem by enforcing $\det[M_u M_d]$ to be real through a non-inverted symmetry \cite{Liang:2025dkm}. However, this solution seems not easy to be embedded in the grand unification, which motivates us in this paper to extend their approach in the framework of GUT-inspired SM. 
\footnote{Recently, several models have been proposed to solve the problem by obtaining the real $\det[M_u M_d]$ based on a discrete group-like symmetry \cite{Antusch:2013rla,Feruglio:2023uof,Petcov:2024vph,Penedo:2024gtb}, where supersymmetry (SUSY) plays an important role. 
Another approach \cite{Liang:2024wbb} uses a six-dimensional spacetime with $\mathbf{T}^2/\mathbb{Z}_3$ orbifold compactification.
}

It is extremely remarkable that the present extension to a $SU(5)$-GUT inspired SM provides us with a very beautiful structure for the non-invertible symmetry. It seems to indicate the presence of some deep fundamental physics at high energies.
$\mathbb{Z}_2$ gauging of $\mathbb{Z}_N$ symmetries, which we use in this paper, can be originated from string compactification, e.g. $T^2/\mathbb{Z}_N$ orbifold compactification with the background magnetic flux $N$ \cite{Kobayashi:2024yqq}.
In particular, we can realize three-generation models by $T^2/\mathbb{Z}_N$ orbifold compactification with the background magnetic flux $N=4$ or 5 \cite{Abe:2008fi}. In fact, we show in this letter that the $N=5$ case provides us with a successful and very attractive non-invertible symmetry in the GUT inspired SM.

We assume, throughout this letter, the CP invariance at the high-energy fundamental level. And we further consider the CP symmetry is spontaneously broken down at some intermediate energy scale. It is extremely important in the LY approach that once the physical angle $\bar\theta$ is vanishing at some high energy scale it remains very small, $\bar\theta\simeq 10^{-16}$, at the weak scale \cite{Ellis:1978hq}.

This paper is organized as follows.
In section \ref{sec:review}, we give a brief review on $\mathbb{Z}_2$ gauging of $\mathbb{Z}_N$ symmetries.
In section \ref{sec:model}, we discuss our model to solve the strong CP problem.
Section \ref{sec:con} is devoted to our conclusions.

\section{Non-invertible selection rules}
\label{sec:review}

In this section, we review non-invertible selection rules realized in $\mathbb{Z}_2$ gauging of $D_N\cong \mathbb{Z}_N\rtimes \mathbb{Z}_2$~\cite{Kobayashi:2024yqq,Kobayashi:2024cvp}. 
The generator of $\mathbb{Z}_N$ symmetry obeys
\begin{align}
    a^N=e,
\end{align}
with $e$ being the identity. Given all of the $\mathbb{Z}_N$ elements $a^k$ ($k=0,1,...,N-1$) (mod $N$), the multiplication rule of two elements $a^{k_1}$ and $a^{k_2}$ obeys the conventional group-like form:
\begin{align}
    a^{k_1} a^{k_2} =a^{k_1+k_2}.
\end{align}
Then, let us focus on the outer automorphism of $\mathbb{Z}_2$ which is generated by the generator $b$ with
\begin{align}
    b^2=e,\qquad ba^k b^{-1}=a^{-k}.
\end{align}
When we gauge this $\mathbb{Z}_2$, one can pick up the $\mathbb{Z}_2$ invariant conjugacy class of $D_N\cong \mathbb{Z}_N\rtimes \mathbb{Z}_2$, i.e.,
\begin{align}
    [g^{(k)}]=\{b^n a^k b^{-n}| n=0,1\} = \{a^k, a^{-k}\},
\end{align}
which includes two elements except for $k=0$ and $k=N/2$ for even $N$. 
We call this procedure as $\mathbb{Z}_2$ gauging of $\mathbb{Z}_N\rtimes \mathbb{Z}_2$ which is represented as $\tilde{\mathbb{Z}}_N$ in what follows. 
Hence, the multiplication rule of two classes is different from the group-like one:
\begin{align}
    [g^{(k_1)}]\otimes [g^{(k_2)}] =[g^{(k_1+k_2)}] + [g^{(k_1-k_2)}],
\end{align}
which is commutable. 

Let us assume that fields in 4D quantum field theory are labeled by a class $[g^{(k)}]$. 
This idea is motivated by string compactifications. 
For instance, through the Kaluza-Klein reduction of higher-dimensional Yang-Mills theory on toroidal backgrounds, 4D massless fields have the global $\mathbb{Z}_N$ flavor symmetry in the context of type IIB magnetized D-brane models~\cite{Abe:2009vi,Berasaluce-Gonzalez:2012abm,Marchesano:2013ega}. When we divide toroidal backgrounds by $\mathbb{Z}_2$ twist, the $\mathbb{Z}_N$ symmetry is broken, but $\mathbb{Z}_2$ invariant modes remaining in the effective action are labeled by a class $[g^{(k)}]$ on the $\mathbb{Z}_2$ orbifold~\cite{Kobayashi:2024yqq}. 
In this case, the couplings of 4D fields $\phi_{k_1}\cdots \phi_{k_n}$ are allowed when the multiplication of corresponding classes $[g^{(k_1)}]\cdots [g^{(k_n)}]$ includes the class $[g^{(k=0)}]$, i.e., 
$\sum_i \pm k_i=0$ (mod $N$). Other string compactifications lead to these non-invertible selection rules in e.g., 4D $E_6$ GUT in the context of heterotic string theory on Calabi-Yau threefolds with standard embedding~\cite{Dong:2025pah}.
Note that 4D field $\phi$ and its conjugate correspond to the same class $[g^{(k)}]$.

In the following, we discuss phenomenological applications of the non-invertible selection rules originating from $\tilde{\mathbb{Z}}_N$\footnote{Other selection rules with and without gaugings are studied in Ref.~\cite{Dong:2025jra}, which will be left for future work.}. In particular, we utilize the case with $N=5$, where there exist three classes:
\begin{align}
    \{[g^0], [g^1], [g^2]\}
\label{eq:tZ5class}
\end{align}
obeying the following fusion rule:
\begin{align}
    [g^0]\otimes [g^0] &= [g^0]\,,\nonumber\\
    [g^0]\otimes [g^1] &= [g^1]\otimes [g^0] = [g^1]\,,\nonumber\\
    [g^0]\otimes [g^2] &= [g^2]\otimes [g^0] = [g^2]\,,\nonumber\\
    [g^1]\otimes [g^1] &= [g^0] + [g^2]\,,\nonumber\\
    [g^1]\otimes [g^2] &= [g^2]\otimes [g^1] = [g^1] + [g^2]\,,\nonumber\\
    [g^2]\otimes [g^2] &= [g^0] + [g^1]\,.
\end{align}

As mentioned in Introduction, $\mathbb{Z}_2$ gauging of $\mathbb{Z}_N$ symmetries can be realized in string compactification such as $T^2/\mathbb{Z}_2$ orbifold compactification with background magnetic flux $N$.
In particular, magnetized orbifold models with $N=5$ and 4 lead to three generations of matter fields \cite{Abe:2008fi}.

Suppose that three generations of both left-handed and right-handed matter correspond to the three classes, $[g^0]$, $[g^1]$, and $[g^2]$.
When the Higgs field corresponds to the class $[g^0]$, one can obtain the following Yukawa matrix:
\begin{align}
\label{eq:Y5-1}
    \begin{pmatrix}
        \checkmark & 0 & 0 \\
        0 & \checkmark & 0 \\
        0 & 0 & \checkmark 
    \end{pmatrix},
\end{align}
where the check symbol $\checkmark$ denotes allowed coupling coefficient.
That is the diagonal matrix.
When the Higgs field corresponds to the class $[g^1]$, one can obtain the following Yukawa matrix:
\begin{align}
\label{eq:Y5-2}
    \begin{pmatrix}
         0 & \checkmark & 0 \\
        \checkmark & 0  & \checkmark\\
        0 & \checkmark & \checkmark 
    \end{pmatrix}.
\end{align}
That is the so-called nearest neighbor interaction form.
When the Higgs field corresponds to the class $[g^2]$, one can obtain the following Yukawa matrix:
\begin{align}
\label{eq:Y5-3}
    \begin{pmatrix}
         0 & 0 & \checkmark \\
        0 & \checkmark   & \checkmark\\
        \checkmark & \checkmark & 0 
    \end{pmatrix}.
\end{align}
These are interesting forms.
We will discuss them in the next section.

\section{Model}
\label{sec:model}

Here, we study a solution to the strong CP problem.
We start with the 4D CP invariant theory, assuming that the CP symmetry is exact in the underlying theory such as string theory and its compactifications.
That is, we assume that the QCD phase $\theta$ is vanishing, and all of Yukawa couplings and other couplings are real in the  high energy fundamental theory. 
However, we have  CP-violating phases in the quark mass matrices at the electroweak energy scale. Thus, spontaneous CP violation should occur within the framework of 4D field theory at an intermediate energy scale. 
Then, the CP phase would appear in quark mass matrices through the Higgs sector.
In general, that induces non-vanishing QCD phase $\bar \theta$ as
\begin{align}
    \bar \theta = \theta + \arg \det[M_u M_d],
\end{align}
where $M_u$ and $M_d$ denote the up- and down-type quark mass matrices, 
even if the bare QCD phase $\theta$ is vanishing. 
In general, the non-vanishing weak CP phase and  
$\arg \det[M_u M_d] = 0$ are not always consistent with each other.
However, those are consistent in special mass matrices.
In Ref.~\cite{Tanimoto:2016rqy}, such textures were studied.
One of them is as follows,
\begin{align}
\label{eq:texture}
    M_u=\begin{pmatrix}
        m_u & 0 & 0\\
        0 & m_c & 0 \\
        0 & 0 & m_t
    \end{pmatrix}, \qquad 
    M_d=\begin{pmatrix}
        0 & a & 0 \\
        a' & be^{i\phi} & c \\
        0 & c' & d
    \end{pmatrix},
\end{align}
where $a,a',b,c,c'$ and $d$ are real, 
while the other textures in Ref.~\cite{Tanimoto:2016rqy} correspond to asymmetric texture zeros in $M_d$.
Here, we focus on the above texture, whose zero entries are symmetric.
Note that the diagonal matrix for $M_u$ can be easily obtained as  Eq.~(\ref{eq:Y5-1}) shown in the previous section.
Also the texture for the above $M_d$ is almost realized as in Eq.~(\ref{eq:Y5-2}) except for the (2,2) entry.
The missing (2,2) entry corresponds the allowed entry in both Eqs.~(\ref{eq:Y5-1}) and (\ref{eq:Y5-3}).

We consider the quark sector as well as the lepton sector together as $SU(5)$ multiplets.
That is, we consider three $SU(5)$ multiplets ${ \bf{\bar{5}},\bf{10}}$ involving the left-handed notations of quarks,  lepton doublets:
\begin{align}
    T &= 
    \begin{pmatrix}
    0 & u_b^c & -u_g^c & -u_r & -d_r\\
    -u_b^c & 0 & u_r^c & -u_g &  -d_g\\    
    u_g^c & -u_r^c  & 0 & -u_b & -d_b\\    
    u_r & u_g & u_b & 0 & e^c\\    
    d_r & d_g & d_b & -e^c & 0\\        
    \end{pmatrix}
    = 
    \begin{pmatrix}
    \epsilon^{\alpha \beta \gamma}\bar{U}_\gamma & -Q^{\alpha s}\\
    Q^{\beta r} & \epsilon^{rs}\bar{E}\\ 
    \end{pmatrix}
,\nonumber\\
    \overline{F}&=
    \begin{pmatrix}
    d_r^c\\
    d_g^c\\
    d_b^c\\
    e\\
    -\nu\\
    \end{pmatrix}
    = 
    \begin{pmatrix}
    \bar{D}_\alpha\\
    \epsilon_{rs}L^s
    \end{pmatrix}
,
\end{align}
where $\{\alpha,\beta, \gamma\}$ and $\{r, s\}$ denote $SU(3)_C$ and $SU(2)_L$ indices, respectively.\footnote{The existence of right-handed neutrinos is not relevant to our discussions.}

We also introduce three Higgs doublet fields $H_1$ and $H_{2,3}$ which belong to the $\bf{\bar{5}}$ representation of the $SU(5)$ GUT. We will not discuss the doublet-triplet splitting problem in this letter. (This is a reason why we call the present model as GUT-inspired SM \cite{Nakayama:2010vs}.) They have the following Yukawa terms:
\begin{align}
    f_u TTH_1^\dagger 
+ f_d\overline{F}TH_2    +f_d^\prime \overline{F}TH_3
  .
    \label{eq:SU5}
\end{align}

In order to realize the texture (\ref{eq:texture}), 
we impose a direct product of two $\tilde{\mathbb{Z}}_5$ symmetries, i.e., $\tilde{\mathbb{Z}}_5^{(1)}\times \tilde{\mathbb{Z}}_5^{(2)}$ in our construction. 
Recalling that there are three classes for each $\tilde{\mathbb{Z}}_5$ shown in Eq.~\eqref{eq:tZ5class}, we consider the following assignments for the three fermion matter fields and for the three Higgs bosons:
\begin{align}
    T&: ([g^0][g^0], [g^1][g^1], [g^2][g^2])\,,\nonumber\\ 
    \overline{F}&: ([g^0][g^0], [g^1][g^1], [g^2][g^2])\,,\nonumber\\ 
    H_1 &: [g^0][g^0]\,,\nonumber\\
    H_2  &: [g^1][g^1]\,,\nonumber\\
    H_3  &: [g^2][g^0]\,.
\end{align}
It leads to the following Yukawa textures:
\begin{align}
    f_u =
    \begin{pmatrix}
        \checkmark & 0 & 0\\
        0 & \checkmark & 0\\
        0 & 0 & \checkmark
    \end{pmatrix}
    \,,\qquad
    f_d =
    \begin{pmatrix}
        0 & \checkmark & 0\\
        \checkmark & 0 & \checkmark\\
        0 & \checkmark & \checkmark
    \end{pmatrix}
    \,,\qquad    
    f_d^\prime =
    \begin{pmatrix}
        0 & 0 & 0\\
        0 & \checkmark & 0\\
        0 & 0 & 0
    \end{pmatrix}
    \,.
\end{align}

To realize these textures, we need to distinguish $H_1$ and $H_{2,3}$.
Otherwise, $H_{2,3}$ are also allowed to couple with $TT$.
Here, we impose the $\mathbb{Z}_3$ symmetry.
We assume that all of $T, \overline{F}, H_{2,3}$,have the $\mathbb{Z}_3$ charge one, while $H_1$ has $\mathbb{Z}_3$ two.
Then, the above mass matrices are realized.
One more interesting point is that since $T$ and $\overline{F}$ have the same charge, we can extend this model to $SO(10)$ GUT. 
The charge assignments are summarized in Table~\ref{tab:charge}.
\begin{table}[H]
    \centering
    \caption{Charge assignments of matter fields under $\tilde{\mathbb{Z}}_5^{(1)}\times \tilde{\mathbb{Z}}_5^{(2)}$ and $\mathbb{Z}_3 \times \mathbb{Z}_2^{(2)}\times \mathbb{Z}_2^{(3)}$ symmetries.}
    \label{tab:charge}
    \begin{tabular}{|c|c|c||c|c|c|c|c|c|}
    \hline
         & $T$ & $\overline{F}$ & $H_1$ & $H_2$ & $H_3$ & $\eta_1$ & $\eta_2$ & $\eta_3$ \\\hline
         $\tilde{\mathbb{Z}}_5^{(1)}$ & ($[g^0]$, $[g^1]$, $[g^2]$) & ($[g^0]$, $[g^1]$, $[g^2]$) & $[g^0]$ & $[g^1]$ & $[g^2]$ & $[g^0]$ & $[g^1]$ & $[g^2]$ \\\hline
         $\tilde{\mathbb{Z}}_5^{(2)}$ & ($[g^0]$, $[g^1]$, $[g^2]$) & ($[g^0]$, $[g^1]$, $[g^2]$) & $[g^0]$ & $[g^1]$ & $[g^0]$ & $[g^0]$ & $[g^1]$ & $[g^1]$ \\\hline
         $\mathbb{Z}_3$ & (1,1,1) & (1,1,1) & 2 & 1 & 1 & 1 & 1 & 0 \\\hline
      $\mathbb{Z}_2^{(2)}$ & even & even & even & even & even & even & odd & even \\\hline 
      $\mathbb{Z}_2^{(3)}$ & even & even & even & even & even & even & even & odd \\\hline 
    \end{tabular}
\end{table}

The potential in the Higgs sector can be written by
\begin{align}
    V_H = &-m_1^2 H_1^\dagger H_1 -m_2^2 H_2^\dagger H_2 -m_3^2 H_3^\dagger H_3 
    \nonumber\\
    &+
    \lambda_1 (H_1^\dagger H_1 )^2 +
    \lambda_2 (H_2^\dagger H_2)^2+
    \lambda_3 (H_3^\dagger H_3)^2
    \nonumber\\
    &+\lambda_{12} H_1^\dagger H_2^\dagger H_1H_2+ \lambda_{13} H_1^\dagger H_3^\dagger H_1H_3+
    \lambda_{23} H_3^\dagger H_2^\dagger H_3H_2 \nonumber \\
    &+ \lambda_{23}'[(H_2^\dagger H_3)^2 +(H_3^\dagger H_2)^2 ],
    \label{eq:potential}
    \end{align}
where all of coupling constants are real.
When $\lambda_{23}'<0$, the CP is not violated spontaneously.

Now, we introduce new scalar bosons, $\eta_i$ for $i=1,2,3$ to generate the spontaneous CP violation.
Then, we have the following potential:
\begin{align}
\label{eq:VHeta}
 V_{H-\eta}=\alpha_{23}(H_2^\dagger H_3 \langle \eta_3 \rangle + \mathrm{h.c.}) + 
 \alpha_{12} (H_1^\dagger H_2 \langle \eta_2 \rangle + \mathrm{h.c.}),
\end{align}
where $\alpha_{12}$ and $\alpha_{23}$ are real dimension one parameters. Notice here that the parameter $\alpha_{12} \simeq m^2/\langle\eta_2\rangle$ and $\alpha_{23} \simeq m^2/\langle\eta_3\rangle$. Here, the $m$ should be less than a few hundred GeV, otherwise the Higgs bosons have too large masses. The parameters $\alpha_{12}$ and $\alpha_{23}$ are indeed extremely small since the vacuum expectation values (VEVs) of $\eta_i$ should be very large like $10^{10} \mathrm{GeV}$. However, their smallness can be naturally explained since we have an enhanced $\mathbb{Z}_2^{(2)} \times \mathbb{Z}_2^{(3)}$ symmetry in the limit where they are vanishing. Under the $\mathbb{Z}_2^{(2)}$ ($\mathbb{Z}_2^{(3)}$) symmetry, the $\eta_{2}$ ($\eta_3$) transforms as odd and all other fields as even, as summarized in Table~\ref{tab:charge}. We impose the $\mathbb{Z}_2^{(2)} \times \mathbb{Z}_2^{(3)}$ symmetry in the $\eta$ boson sector to control their potential as shown below.

If the vacuum expectation values $\langle \eta_i \rangle$ violate the CP symmetry, such a CP violation induces the CP violation in the Higgs sector.
To allow the above couplings in Eq.~(\ref{eq:VHeta}), the classes of $\eta_i$ must be assigned as 
\begin{align}
   & \eta_2: [g^1][g^1],\qquad
     \eta_3: [g^2][g^1],
\end{align}
and $\eta_2$ and $\eta_3$ have $\mathbb{Z}_3$ charges, 1 and 0, respectively.
In addition, we assign the classes of $\eta_1$ as follows,
\begin{align}
    \eta_1:[g^0][g^0],
\end{align}
and it has the $\mathbb{Z}_3$ charge one.
In addition to $V_{H-\eta}$, we have the following terms:
\begin{align}
    -\sum_{i,j=1}^3 c_{ij}|\langle \eta_j\rangle|^2 H^\dagger_iH_i.
\end{align}
These are additional contributions to $m_i^2$.
We renormalize them to $m_i^2$.

The potential of $\eta_i$ can be given by
\begin{align}
\label{eq:Veta}
    V_\eta = & -\mu_3 \eta_3 \eta_3^\dagger-\mu_3'(\eta^2_3+\eta_3^{\dagger 2})
    +\xi_3 (\eta_3 \eta_3^\dagger)^2+\xi'_3(\eta_3^4 + \eta_3^{\dagger 4})+ \xi'_{33}(\eta_3^2 + \eta_3^{\dagger 2})\eta_3\eta_3^\dagger \nonumber  \\
    &-\mu_1 \eta_1 \eta_1^\dagger-\mu_2\eta_2 \eta_2^\dagger +\xi_1 (\eta_1\eta_1^{\dagger})^2  +\xi_2 (\eta_2\eta_2^{\dagger})^2  +\xi_{12} (\eta_1\eta_1^{\dagger})(\eta_2\eta_2^{\dagger})\nonumber \\
    & + \sum_{i=1,2} \left[ \xi_{3i}  \eta_3 \eta_3^\dagger  +
    \xi'_{3i} (\eta^2_3+\eta_3^{\dagger 2})\right]\eta_i \eta_i^\dagger \nonumber \\
    &-\beta (\eta_1^3 + \eta_1^{\dagger 3}) - \beta'(\eta_1 \eta_2^2 + \eta_1^{\dagger} \eta_2^{\dagger 2}) 
    -\xi_{12}'\left[ (\eta_1^\dagger \eta_2 )^2+ (\eta_1 \eta_2^\dagger)^2 \right],
\end{align}
where all of the coefficients are real. Here, we have imposed a new symmetry $\mathbb{Z}_2^{(2)}\times \mathbb{Z}_2^{(3)}$. We remark that $\eta_{2}$ ($\eta_3$) transforms as odd and all other fields as even under the $\mathbb{Z}_2^{(2)}$ ($\mathbb{Z}_2^{(3)}$) symmetry. This is the symmetry that can explain naturally the smallness of the parameters $\alpha_{12}$ and $\alpha_{23}$ in Eq.~(\ref{eq:VHeta}). Notice that this $\mathbb{Z}_2^{(2)}\times \mathbb{Z}_2^{(3)}$  symmetry is approximate one in the total Lagrangian, but we consider it is exact in the $\eta$ sector at the tree level.

The potential terms in the first line include only $\eta_3$.
This potential was studied in Refs.~\cite{Liang:2024wbb,Liang:2025dkm} and it was shown that the CP symmetry is spontaneously broken. (The phase depends also on the VEVs of $\eta_i$ due to the couplings in the third line.)
That is, the vacuum expectation value $\langle \eta_3 \rangle$ has non-vanishing phase.
The above potential has an additional term with the coefficients $\xi_{3i}'$.
We denote $\eta_i=|\eta_i|e^{i\varphi_i}$.
Then, the potential terms relevant to the phase $\varphi_3$ are written by 
\begin{align}
 V_{3\varphi_3} &=  -2\mu'_3|\eta_3|^2\cos 2\varphi_3 + 2\xi_3'|\eta_3|^4\cos 4\varphi_3 +2|\eta_3|^2 \sum_{i=1,2,3}\xi_{3i}'|\eta_i|^2\cos 2\varphi_3 \nonumber \\
 &= -2|\eta_3|^2\cos 2\varphi_3 \left( \mu_3' -\sum_{i=1,2,3}\xi_{3i}'|\eta_i|^2 \right) +2\xi'_3|\eta_3|^4 (2\cos^2 2\varphi_3 -1).
\end{align}
Similar to Refs.~\cite{Liang:2024wbb,Liang:2025dkm}, 
the potential minimum corresponds to non-vanishing phase $\varphi_3 \neq 0, \pi$ and $\cos 2\varphi \neq 1$ for generic parameters.

The phases of $\eta_1$ and $\eta_2$ appear only in the last two terms of $V_\eta$.
These terms are written by 
\begin{align}
    -2\beta |\eta_1|^3 \cos 3\varphi_1-2\beta'|\eta_1||\eta_2|^2\cos(\varphi_1+2\varphi_2) 
    -2\xi_{12}'|\eta_1|^2|\eta_2|^2 \cos(2\varphi_1-2\varphi_2).
\end{align}
When all of $\beta$, $\beta'$, and  $\xi_{12}'$ are positive, the potential minimum corresponds to 
\begin{align}
    \varphi_1=\varphi_2=\frac{2\pi m}{3},
\end{align}
with $m=0,1,2$.

Now let us study the CP phases of the Higgs vacuum expectation values $H_i$, which are denoted by $|H_i|e^{i\phi_i}$.
The potential terms in $V_{H-\eta}$ are important.
We use the $U(1)$-hypercharge rotation so that $\langle H_2 \rangle$ is real.
Since $\langle \eta_3 \rangle$ has non-vanishing phase, 
the VEV $\langle H_3 \rangle$ must have non-vanishing phase because of the 
term with the coefficient $\alpha_{23}$ in $V_{H-\eta}$. 
Indeed, the relevant term can be written by 
\begin{align}
    2\alpha_{23}|H_2||H_3|\cos(\varphi_3+\phi_3).
\end{align}
For example, when $\alpha_{23} <0$, the potential minimum corresponds to 
$\phi_3=-\varphi_3$, where the value of $\varphi_3$ depends on parameters.

Next, we study the phase of $\langle H_1 \rangle$ and we write its phase by $\phi_1$.
The term with the coefficient $\alpha_{12}$ in $V_{H-\eta}$ is written by
\begin{align}
    2\alpha_{12}|H_2||H_1|\cos(\varphi_2-\phi_1).
\end{align}
Suppose that $\alpha_{12} <0$.
Then, the potential minimum corresponds to 
\begin{align}
    \phi_1=\varphi_2=\frac{2\pi m}{3}.
\end{align}

Obviously, if $\phi_1=0$, the quark masses in the up sector are real.
Even if the phase is non-vanishing, we find 
the phase factor of $\det M_u$ is $e^{2\pi mi}$ because $3 \times 3$ matrix.
As a result, we can realize the texture of Eq.~(\ref{eq:texture}) with 
$\arg \det[M_u]=0$ and non-vanishing $\phi=\phi_3$ in the (2,2) entry of $M_d$.

Now we are at the point to discuss quantum corrections of the above argument. It is known that quantum loop corrections violate the non-invertible symmetry selection rule. (See e.g. Refs.~\cite{Heckman:2024obe,Kaidi:2024wio,Funakoshi:2024uvy}.) In fact, one loop diagrams with intermediate $H_2$ bosons induce unwanted operators, $H_1^\dagger H_3 \eta_2^\dagger \eta_3$ and $H_1^\dagger H_3 \eta_2^\dagger \eta_3^\dagger$, which shift the phase of $H_1$. However, the coefficients are proportional to $\alpha_{12}\alpha_{23}\xi_{32}/m_\eta^2$ for $H_1^\dagger H_3 \eta_2^\dagger \eta_3$ and $\alpha_{12}\alpha_{23}\xi'_{32}/m_\eta^2$ for $H_1^\dagger H_3 \eta_2^\dagger \eta_3^\dagger$ appearing in Eqs.~\eqref{eq:VHeta} and \eqref{eq:Veta}, where $m_\eta$ denotes a (typical) mass scale of the $\eta_i$ sector, up to a loop factor.
The corrections must be small compared with Eq.~(\ref{eq:VHeta}) enough to realize the experimental constraint $\bar \theta < 10^{-10}$.
That requires the ratio $\alpha_{23}\xi_{23}\langle \eta_3\rangle/m_\eta^2 < 10^{-10}$.
For example, when $\xi_{23}\sim 1$, $\alpha_{23} \simeq (100 {\rm GeV})^2/\langle\eta_3\rangle$, and $\langle \eta_3\rangle \sim m_\eta$, 
we need $\langle\eta_3\rangle \sim m_\eta \gg 10^{6}$ GeV.

There are other correction terms due to loop effects. However, if we use couplings between the $\eta_{2,3}$ and the Higgs $H_{1,2,3}$ in Eq.~\eqref{eq:VHeta}, the loop amplitudes are always suppressed by their small coupling constants $\alpha_{ij}$. Therefore, we consider only the interactions in Eq.~\eqref{eq:Veta} which has the $\mathbb{Z}_2^{(3)}$ symmetry, where $\eta_3$ is $\mathbb{Z}_2^{(3)}$ odd and $\eta_1$ and $\eta_2$ are $\mathbb{Z}_2^{(3)}$ even. 
Furthermore, we have an accidental exchange symmetry between $\eta_3$ and $\eta_3^\dagger$. Together with both symmetries, we conclude loop-induced effective operators that are relevant for the phase determination of the $\eta_1$ is $(\eta_1^3 +\eta_1^{\dagger 3})(\eta_3^2+\eta_3^{\dagger  2})$. It is remarkable that the phase of the $\eta_1$ does not shift from $2n\pi/3$ even if the $\eta_3$ has an arbitrary non-vanishing phase $\varphi_3$.

We have not discussed, in this letter, the possible cut-off $\Lambda$ suppressed operators such as $(1/\Lambda)\eta_1^3 \eta_3^2 + \mathrm{h.c.}$, since they depend largely on the physics at the cut-off scale. However, if $|\langle\eta_i\rangle|/\Lambda < 10^{-10}$, the experimental constraint on the physical vacuum angle $\bar\theta$ is satisfied.


\section{Conclusions}
\label{sec:con}

We have constructed a 3 zero texture for the down-type quark mass matrix $M_d$ and a real diagonal texture for the up-type quark mass matrix, based on a non-invertible symmetry, $\tilde{\mathbb{Z}}_5^{(1)}\times \tilde{\mathbb{Z}}_5^{(2)}$, in the $SU(5)$-GUT inspired SM. Together with the assumption of CP invariant in the high-energy fundamental theory, we solve the strong CP problem in QCD \cite{Liang:2025dkm}, since $\det[M_u M_d]$ is real.  The mass texture we have constructed in this letter is exactly the same as one of the mass matrices $M_d^{(1)}$ in \cite{Tanimoto:2016rqy} which is consistent with all of the CKM phenomenology. 

We have started with the CP invariant 4D theory, and  
the CP symmetry is assumed to be spontaneously broken down at some intermediate energy scale.
Such CP violation appears in the Higgs potential and one of the Higgs fields acquires a complex VEV generating the complex CKM matrix. However,
we need a proper texture such that the CP violation does not appear in the QCD phase $\bar \theta$. We have constructed a consistent texture for the quark mass matrices using the non-invertible coupling selection rules derived from $\mathbb{Z}_2 $ gauging of $\mathbb{Z}_5$ symmetries.\footnote{As pointed out in Ref.~\cite{Liang:2025dkm}, quantum corrections to the Yukawa coupling matrices can generate a complex phase in det$[M_dM_u]$ at the two-loop level, but the resulting shift in the vacuum angle, $\bar\theta$, is well below the current experimental bound $\bar\theta <10^{-10}$.}

The fermion matter sector in our model corresponds exactly to $SU(5)$ multiplets \cite{Nakayama:2010vs}, and they can be embedded into $SO(10)$ multiplets, because $\bf 10$ and $\bf \bar 5$ multiplets are assigned to have the same charges of the non-inverted symmetry.
Our model includes three Higgs fields $H_i~(i=1-3)$ and extra scalar fields $\eta_i~(i=1-3)$ as well as three generations of matter fields.
That is, all of the mode numbers are equal to three.
Furthermore, their structures of the first $\tilde{\mathbb{Z}}_5$ class are the same, i.e., $([g^0],[g^1],[g^2])$, for the fermion matters. 
It would be interesting to study the unification of the matter sector, the Higgs sector, and the $\eta$ sector.\footnote{In general, string compactification leads to several Higgs fields. 
In addition, the matter sector and Higgs sector are combined to one sector, e.g. $E_6$ {\bf 27} multiplets in heterotic string theory with standard embedding.}
However, it may be very difficult to generate the desired breaking of such a unification model down to the SM.

The three doublet Higgs bosons are one of crucial predictions at the weak scale in the present model. In particular, two Higgs bosons couple to the down-type quark and charged lepton sectors and hence it causes the flavor changing neutral current events in the down-type quark and the charged lepton sector. This is a smoking gun in the present model.

We have not studied details in the lepton sector in this letter. However, it is a straightforward task to include the lepton sector if we introduce three right-handed neutrinos, $N_{Ri}~(i=1-3)$. We will construct the model and discuss the CP violation in the neutrino oscillation and the baryon-number creation through the leptogenesis \cite{Fukugita:1986hr} in a future publication \cite{Kobayashi:XXX}.

\acknowledgments

The authors would like to thank Qiuyue Liang  for useful discussions.
This work was supported by JSPS KAKENHI Grant Numbers JP23K03375 (T.K.), JP25H01539 (H.O.), and JP24H02244 (T.T.Y.). T.~T.~Y.~was supported also by the Natural Science Foundation of China (NSFC) under Grant No.~12175134 as well as by World Premier International Research Center Initiative (WPI Initiative), MEXT, Japan.

\bibliography{references}{}
\bibliographystyle{JHEP}

\end{document}